# Detection and imaging of chemicals and hidden explosives using terahertz time-domain spectroscopy and deep learning

Xinghe Jiang[1, 2, +], Yuhang Li[1, 2, 3, +], Yuzhu Li[1, 2, 3], Che-Yung Shen[1, 2, 3], Aydogan Ozcan[1, 2, 3, *] & Mona Jarrahi[1, 2, *]

[1]Electrical and Computer Engineering Department, University of California, Los Angeles, California 90095, USA

[2]California NanoSystems Institute (CNSI), University of California, Los Angeles, California 90095, USA

[3]Bioengineering Department, University of California, Los Angeles, California 90095, USA

[+]Equal contributing authors

*Correspondence: Mona Jarrahi and Aydogan Ozcan

**Abstract**

Detecting concealed chemicals and explosives remains a critical challenge in global security. Terahertz time-domain spectroscopy (THz-TDS) offers a promising non-invasive and stand-off detection technique owing to its ability to penetrate optically opaque materials without causing ionization damage. While many chemicals exhibit distinct spectral features in the terahertz range, conventional terahertz-based detection methods often struggle in real-world environments, where variations in sample geometry, thickness, and packaging can lead to inconsistent spectral responses. In this study, we present a chemical imaging system that integrates THz-TDS with deep learning to enable accurate pixel-level identification and classification of different explosives. Operating in reflection mode and enhanced with plasmonic nanoantenna arrays, our THz-TDS system achieves a peak dynamic range of 96 dB and a detection bandwidth of 4.5 THz, supporting practical, stand-off operation. By analyzing individual time-domain pulses with deep neural networks, the system exhibits strong resilience to environmental variations and sample inconsistencies. Blind testing across eight chemicals—including pharmaceutical excipients and explosive compounds—resulted in an average classification accuracy of 99.42% at the pixel level. Notably, the system maintained an average accuracy of 88.83% when detecting explosives concealed under opaque paper coverings, demonstrating its robust generalization capability. These results highlight the potential of combining advanced terahertz spectroscopy with neural networks for highly sensitive and specific chemical and explosive detection in diverse and operationally relevant scenarios.

**Introduction**

The terahertz region of the electromagnetic spectrum, spanning frequencies from approximately 0.1 to 10 THz, bridges the gap between microwave and far-infrared radiation. This spectral range is particularly attractive due to its unique interactions with a variety of materials. The photon energy of terahertz waves corresponds to the excitation energies of numerous molecular rotational and vibrational transitions, while remaining low enough to avoid ionization damage to samples. As a result, terahertz spectroscopy has emerged as a powerful tool across diverse fields, including biomedical diagnosis[1-3], security screening[4-14], pharmaceutical quality control[15-18], as well as agriculture and food inspection[19-22]. The ability of terahertz radiation to penetrate many optically opaque materials—such as paper, clothing, and plastics—makes it especially valuable for the stand-off detection of concealed chemicals. This non-invasive imaging capability enables the identification of hidden threats, such as explosives, without direct contact, thereby improving safety and operational efficiency.

Several technologies are currently employed for explosive detection, including X-ray imaging[23,24], millimeter-wave imaging[25], mass spectrometry[26], ion mobility spectrometry[27,28], and hybrid approaches that integrate multiple modalities. While these systems are critical for enhancing security and mitigating threats, each has specific limitations. For example, X-ray scanners and millimeter-wave imagers can indicate the presence of suspicious objects but often lack chemical specificity to confirm the presence of explosives[29]. Conversely, chemical detection techniques such as mass spectrometry, ion mobility spectrometry, and trained canine units require close proximity to the target, which is often impractical in dynamic or crowded environments. Furthermore, these approaches struggle to detect low-volatility explosives or those embedded in complex matrices, complicating reliable identification.

In light of these limitations, terahertz spectroscopy presents a promising alternative for accurate, stand-off detection of explosives. Of particular interest is terahertz time-domain spectroscopy (THz-TDS), which captures the time-resolved electric field of terahertz radiation, allowing for comprehensive material characterization over a broad spectral range. Numerous studies have shown that many explosives exhibit unique spectral fingerprints in the terahertz domain[8,12,30-32], permitting their identification through distinct absorption features. Statistical techniques such as principal component analysis (PCA) have been employed to analyze terahertz absorbance spectra for the detection of explosives across varying concentrations[33,34], and spectral images have been generated by mapping intensity at selected frequencies[34-36]. In addition, neural networks have been used to enhance target identification accuracy based on terahertz absorption features[31,37]. Despite these advancements, spectral methods relying on the terahertz power spectrum miss the spectral phase information. Thus, they remain susceptible to artifacts from multiple reflections, scattering, and environmental interferences. These effects can distort spectral signatures, even for identical materials. Puc et al.[9] studied these challenges by covering explosive chemicals with paper and textiles and found that secondary reflections significantly altered the observed absorption peaks. To address this, time-domain analysis of the terahertz signal has been proposed as a way to mitigate such environmental artifacts. This approach enables the extraction of robust time-domain features for resolving explosive signatures, as demonstrated in recent studies[38-40]. However, existing time-domain techniques have not yet achieved pixel-level chemical imaging and classification, which is an important capability to reveal the shape and position of the chemicals in the field of view. Therefore, while substantial progress has been made so far, a robust terahertz imaging system capable of accurate, stand-off explosive detection—independent of sample geometry, size, packaging, or environmental variability—remains a critical unmet need.

To address these challenges and detect concealed explosives, we developed a chemical imaging system that integrates THz-TDS with advanced deep learning techniques. This system enables accurate pixel-level identification of hidden explosives and operates in reflection mode, making it particularly well-suited for stand-off detection in real-world scenarios. A key enabler in the developed THz-TDS system is the use of plasmonic nanoantenna arrays for terahertz generation and detection[41-51], which significantly enhance the signal-to-noise ratio compared to the state-of-the-art THz-TDS systems while providing a broad bandwidth (see Supplementary Fig. S1 and Table S1). Operating in reflection mode, the THz-TDS system achieves a peak dynamic range of 96 dB and a detection bandwidth of 4.5 THz in 3 s. The terahertz beam is focused onto samples placed on a scanning stage, which is controlled by an XY motorized translation platform used for raster scanning. During the sample interrogation, the system collects multiple reflected pulses from different internal surfaces, analyzing each pulse individually, regardless of their time sequence. This pulse-resolved approach enables chemical identification at the pixel level that is robust to variations in sample geometry, size, and packaging. Compared to conventional methods that rely solely on power spectra, our approach analyzes terahertz time-domain pulses, capturing both amplitude and phase information. This comprehensive analysis enables more accurate identification of chemicals, including hidden explosives. The acquired temporal waveforms for each pixel of a sample field-of-view (FOV) are processed using deep learning models, including convolutional neural networks (CNNs) and transformer architectures, to automatically classify the chemical content of each pixel.

We experimentally demonstrated that our system was capable of detecting eight chemical species, including four pharmaceutical compounds—microcrystalline cellulose (MCC), dibasic calcium phosphate (DCP), mannitol (MAN), and ibuprofen (IBU)—and four explosives—potassium nitrate ($KNO_3$), pentaerythritol tetranitrate (PETN), cyclotrimethylene trinitramine (RDX), and trinitrotoluene (TNT). In blind testing experiments, the system achieved an average classification accuracy of 99.42% at the pixel level. To further assess the generalization capability of our system, we tested the detection of explosives concealed under paper coverings, yielding an average accuracy of 88.83%. By leveraging deep learning and advanced terahertz spectroscopic imaging, this system offers a highly sensitive and specific platform for rapid, stand-off chemical imaging and target detection. It holds strong potential for applications across pharmaceutical manufacturing, security screening, and industrial quality assurance.

**Results**

A schematic diagram and the operational principles of the terahertz time-domain spectroscopic imaging system are illustrated in Fig. 1a. The system is driven by a Ti:sapphire laser that generates 135 fs pulses at a central wavelength of 800 nm with a repetition rate of 76 MHz. The laser beam is split into two paths to independently pump and probe the photoconductive terahertz source and detector, both based on plasmonic nanoantenna arrays. Under an applied bias voltage, photo-generated carriers in the active area of the terahertz source drift toward the nanoantenna electrodes, producing sub-picosecond terahertz pulses[51]. At the detector, photo-generated carriers drift in response to the incident terahertz field, generating a photocurrent proportional to the field strength[49]. The plasmonic nanoantenna arrays enhance optical absorption near the electrodes through surface plasmon excitation, preserving high quantum efficiency and ultrafast response. This leads to efficient terahertz generation and highly sensitive detection. A mechanical optical delay line in one of the paths introduces a tunable time delay between the pump and probe beams. Terahertz radiation generated at the source is collimated using an off-axis parabolic mirror and then focused onto the sample using a second off-axis parabolic mirror, producing a beam spot with a 2-mm focal size. The sample is positioned on a metallic platform controlled by an XY motorized stage for spatial scanning. The reflected terahertz signal is directed to the detector via a terahertz beam splitter and subsequently focused onto the detector using another off-axis parabolic mirror. The time-domain terahertz electric field at each spatial point of the sample FOV is recorded by measuring the photocurrent from the terahertz detector as a function of the time delay introduced by the optical delay line. Applying a Fourier transform to the time-domain signal yields a broadband power spectrum extending up to 4.5 THz, with a peak dynamic range of 96 dB. The performance of the system in transmission mode is also evaluated (see Supplementary Fig. S1). Each spectrum is acquired within an integration time of 3 seconds using pump and probe powers of 600 mW and 150 mW, respectively. The full time-domain response of the sample is obtained by raster scanning the sample FOV and acquiring 10,100 time-domain points at each spatial location.

Using this terahertz spectroscopic imaging system, we analyzed a range of pharmaceutical and explosive materials. The four pharmaceutical compounds—MCC, DCP, MAN, and IBU—were compressed into cylindrical tablets using a hydraulic pellet press. Each tablet had a diameter of 10 mm, with thicknesses ranging from 2 to 4 mm. The four explosives examined—$KNO_3$, PETN, RDX, and TNT—were not suitable for high-pressure tablet formation due to explosion risk. Instead, they were mounted in an aluminum holder featuring a cylindrical cavity with a 10 mm diameter. The explosive samples placed in the holder had thicknesses between 2 and 4 mm. All samples were raster scanned over a 12 × 12 mm² area with a spatial step size of 1 mm. A 12 × 12 mm² area was scanned in ~ 447 seconds. This duration includes a 3-second acquisition time per pixel plus the time needed for stage movements between successive points.

Following data acquisition, the terahertz signals were processed using two dedicated neural networks—which we termed EdgeNet and ClassNet—for chemical identification and classification. Figure 1b illustrates the full data processing pipeline, encompassing the terahertz imaging, neural network inference, and spatial post-processing. First, the system acquires time-domain waveforms corresponding to each chemical sample. These pulses are then extracted and used as inputs to the two neural networks for edge detection and chemical classification.

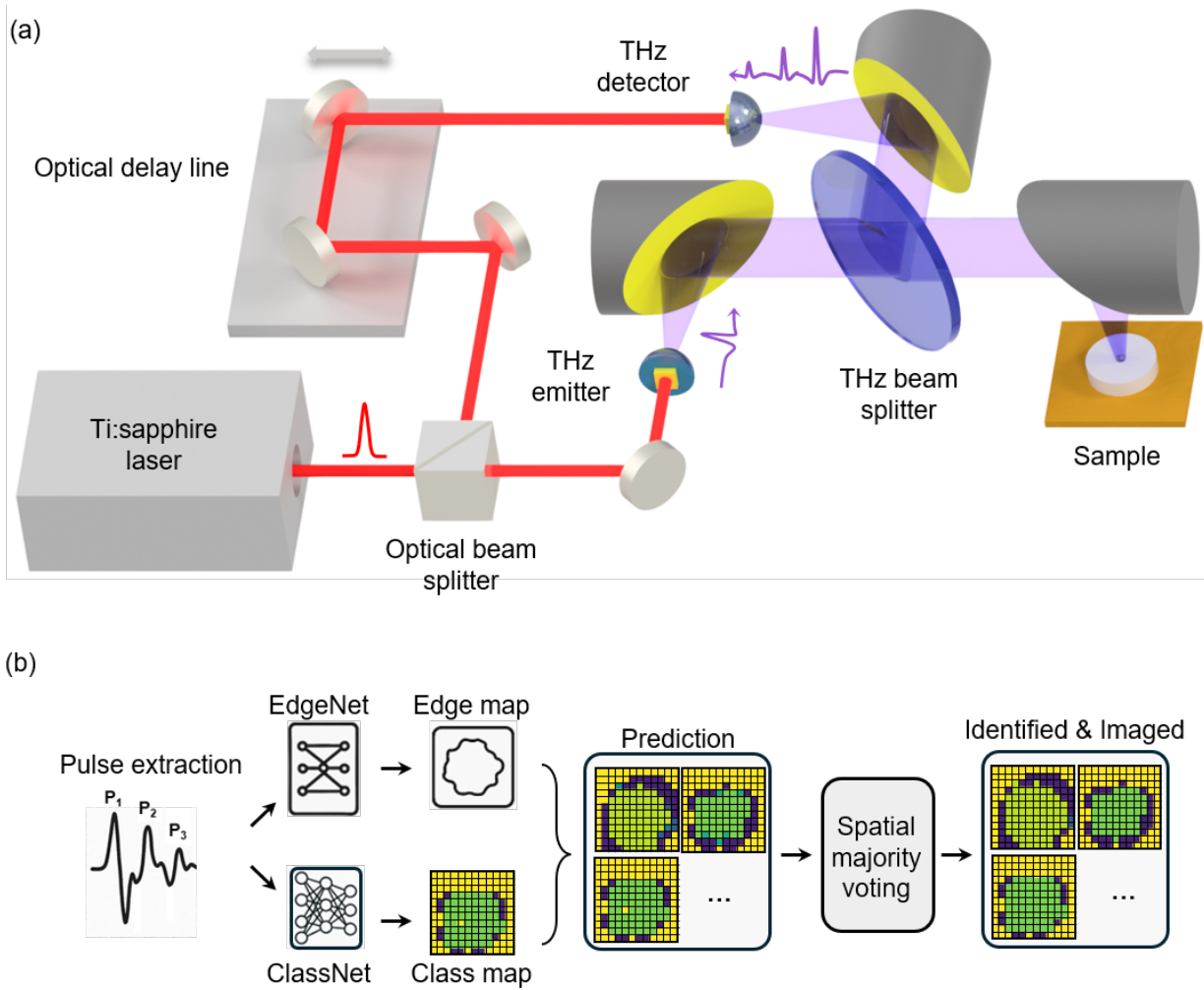

**Figure 1**. (a) Schematic diagram of the terahertz time-domain spectroscopy-based chemical imaging system. The sample is mounted on a metallic stage controlled by a motorized translation stage for raster scanning. (b) Processing pipeline for chemical classification using terahertz spectral imaging. The workflow incorporates two neural networks: EdgeNet, which identifies sample boundaries/edges, and ClassNet, which performs chemical classification per pixel, generating a class map that preserves the spatial resolution of the edge map. Spatial post-processing in the form of pixel-level majority voting is also applied to enhance the classification accuracy of each chemical image.

EdgeNet functions as a binary classifier, distinguishing boundary pixels from non-boundary pixels to identify the presence or absence of edges within the sample FOV. At each pixel location, EdgeNet processes all the extracted terahertz pulses together to make its decisions. In parallel, ClassNet operates on an individual pulse basis, processing each pulse separately to generate probabilistic predictions for the chemical class, including the identification of the background sample holder (metal) if there is no sample for that pixel. For each pixel, these predictions are aggregated using a winner-takes-all strategy, where each pixel is assigned to the class with the highest predicted probability. Finally, the outputs of EdgeNet and ClassNet are integrated through a decision algorithm to determine the final classification of each pixel across the entire sample FOV. To further enhance our image-based classification accuracy, spatial majority voting and morphological operations are applied as post-processing steps (refer to Methods for details). These techniques incorporate local spatial context, helping to suppress inference artifacts and generate cleaner, more accurate chemical images that preserve the true spatial distribution of the material under test.

The distinct input structures for EdgeNet and ClassNet are custom-tailored to their specific objectives. EdgeNet leverages the collective information from all the extracted terahertz pulses to make its decisions. This allows the network to identify complex temporal signatures, such as multiple reflections and scattering artifacts, that are characteristic of material boundaries. In contrast, for chemical classification, ClassNet processes each pulse individually, generating probabilistic predictions for the chemical class of each pixel, including the identification of the background sample holder. This approach was designed to preserve the unique spectral features within each reflection that are critical for identifying the chemical type of interest.

When the terahertz beam is focused onto a sample, as illustrated in Fig. 2a, multiple reflected pulses are captured in a single time-domain scan. Typically, four distinct pulses are observed in this reflection-mode setup, denoted as $P_1$, $P_2$, $P_3$, and $P_4$: $P_1$ corresponds to the reflection from the sample's top surface and contains surface reflectance information; $P_2$ is the reflection from the metal substrate and does not carry chemical information; $P_3$ arises from the sample's bottom interface and contains key absorption features representative of the material; $P_4$ and subsequent pulses are due to internal multiple reflections and also encode absorption characteristics of the sample under test. For reference, the pulse reflected from the metal background is nearly identical in shape to $P_2$ (see Supplementary Fig. S2). While our analysis focuses on these four primary reflections, a time-domain trace can contain additional pulses. These typically arise from artifacts like edge scattering or from extra layers like a paper cover, which we will detail in subsequent measurements. The relative sequence and timing of these pulses vary depending on how the sample is positioned, and this information is assumed to be unknown to our neural networks.

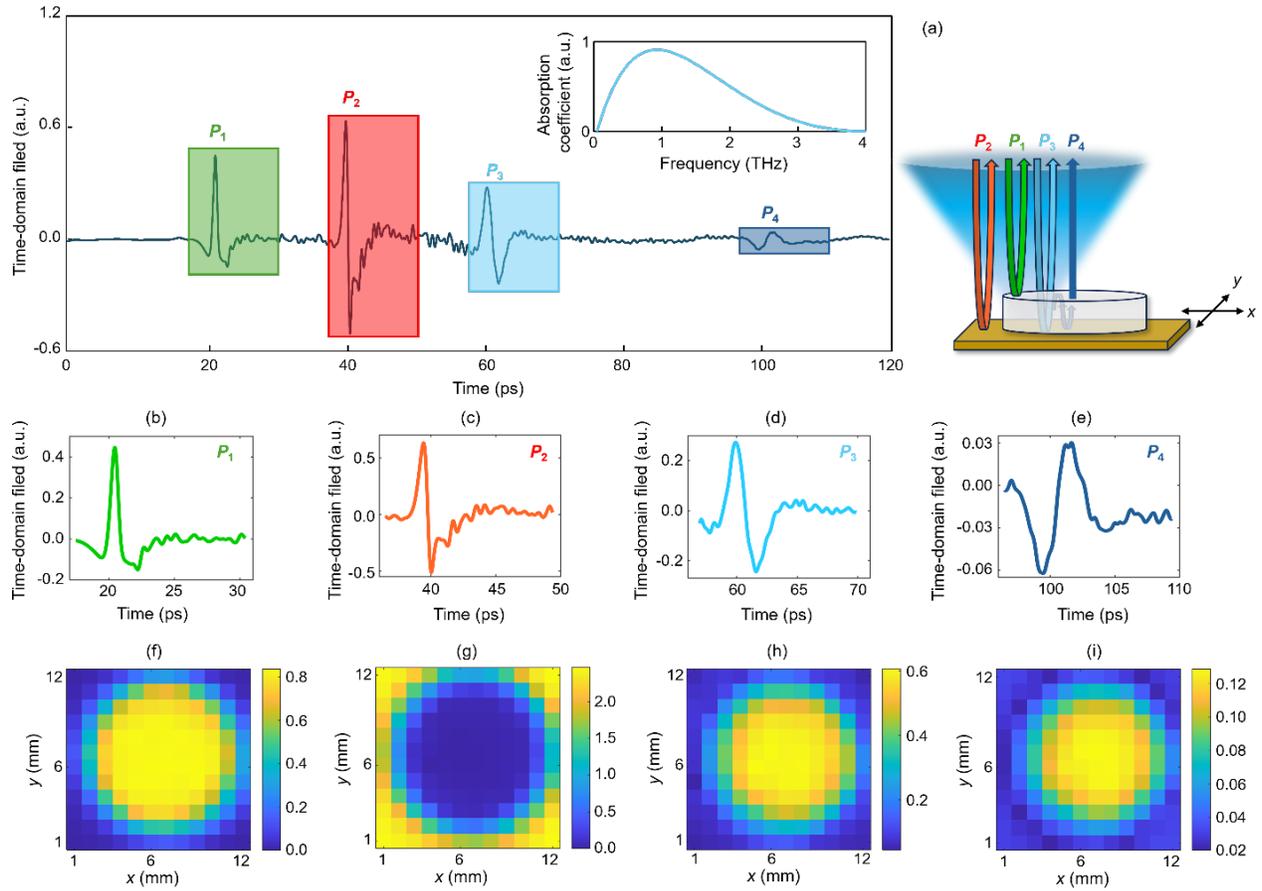

**Figure 2**. (a) Reflected terahertz time-domain response from a KNO$_3$ sample. When the terahertz beam is incident near the sample edges, at least four distinct reflected pulses are observed. The first pulse (P$_1$) corresponds to the reflection from the sample's top surface; the second pulse (P$_2$) originates from the metallic background (sample holder); the third pulse (P$_3$) arises from the bottom interface of the sample; and the fourth one (P$_4$) results from an internal reflection following two round trips within the sample. The inset shows the calculated terahertz absorption spectrum of KNO$_3$, derived from the third and second pulses. (b–e) Cropped waveforms of the four reflected pulses by pulse extraction algorithm in (a), each with a duration of 13 ps. After performing pulse extraction for every pixel across the 12 mm × 12 mm raster scan area, (f–i) represent the color maps of the peak amplitude values for each pulse (P$_1$-P$_4$).

While the material's terahertz absorption spectrum can be obtained by calculating the Fourier transform of $P_3$ relative to the reference pulse $P_2$, as illustrated in the inset of Fig. 2a, directly applying a Fourier transform to the entire time-domain waveform can be unreliable, as interfering pulses may distort the spectral signatures. To mitigate this, we adopt a pulse-based chemical classification strategy. This approach significantly improves the robustness of inference by enabling classification that is insensitive to pulse order or temporal spacing—parameters that can fluctuate with changes in the sample geometry, placement, and packaging. Each full time-domain waveform per pixel is segmented into individual pulses with a duration of 13 ps, each consisting of 650 data points, as illustrated in Figs. 2b-2e. The 13-ps time window is sufficient to capture a terahertz pulse and provides a spectral resolution of 76.9 GHz, which is adequate for our classification task. After performing the same pulse extraction for every pixel across the FOV, we take the peak amplitude value for each pulse ($P_1$-$P_4$) at each pixel to construct the 2D maps, as shown in Figs. 2f-2i. The peak amplitudes of these pulses exhibit strong spatial correlation and track the geometry of the sample. This peak amplitude data is also central to our automated processing pipeline. For pulse extraction, a global detection threshold is applied to all waveforms to consistently identify potential pulses while rejecting noise. Following the pulse extraction step, the algorithm performs two labeling tasks for generating the training data. The first step of this training label generation categorizes each pulse—distinguishing between sample and background reflections—by using pulse timing in conjunction with prior knowledge of the sample's type and structure. The algorithm then employs a sample-specific amplitude threshold, which is determined by analyzing peak amplitude distributions across various samples, to identify the edges. Any pixel where all the detected pulse amplitudes fall below this second threshold is labeled as an "edge" (see Methods for details). These isolated pulses, without their relative time delay information, are used as individual inputs to a neural network trained to predict the chemical composition at each spatial location. This strategy enhances the system's resilience to environmental variability and enables reliable, pixel-level chemical identification across diverse real-world scenarios, including concealed chemicals.

Figure 3 illustrates the architectures of the two neural networks used in our chemical imaging and classification pipeline: EdgeNet and ClassNet. EdgeNet (Fig. 3a) is designed for binary edge detection and is implemented using a 4-layer transformer encoder[52]. It accepts an input tensor of size 8×650, where each row corresponds to a distinct terahertz pulse; if less than 8 pulses are available per pixel, zeros are padded for the remaining pulse entries. Self-attention-based feature extraction is applied to extract features across the pulse set[53], followed by a mean pooling operation along the sequence dimension. The resulting feature vector is passed through a fully connected layer to generate class probabilities indicating the presence or absence of an edge per pixel within the sample FOV. Notably, we did not incorporate temporal positional encoding, which is commonly used in transformer architectures, as the order of the input pulses may be unknown or not repeatable in real-world scenarios (e.g., for concealed chemicals). This design choice improves the model's robustness to unknown variations in pulse ordering. The internal structure of the transformer encoder (Fig. 3b) consists of layer normalization, a multi-head self-attention mechanism, and a feedforward multilayer perceptron (MLP), connected through residual pathways to facilitate gradient flow. The multi-head attention module (Fig. 3c) comprises eight parallel attention heads (h = 8), each computing scaled dot-product attention over the input sequence.

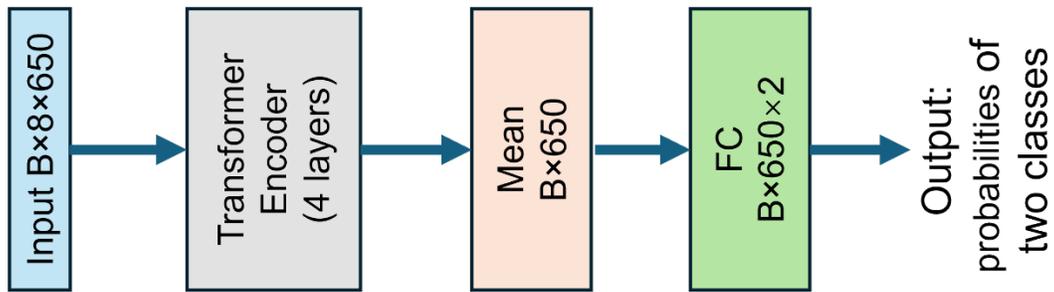

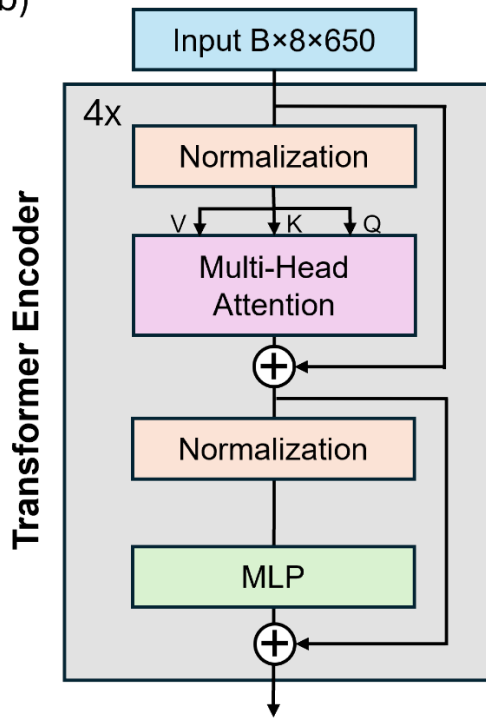

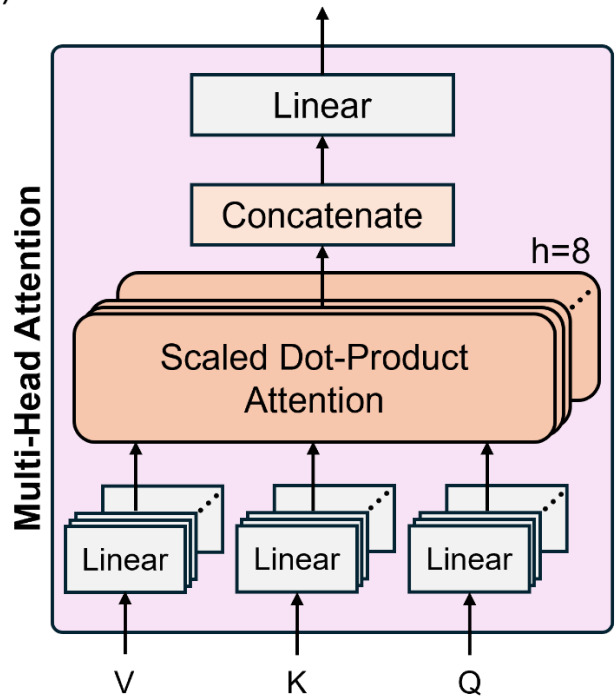

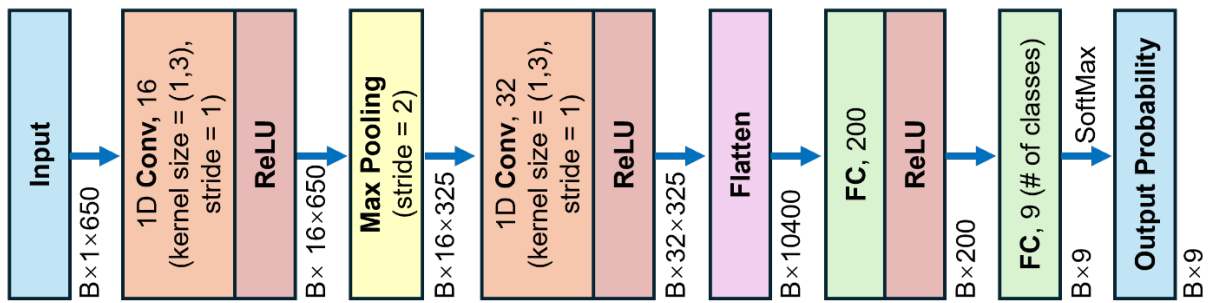

**Figure 3**. Network architectures used in the chemical detection and classification pipeline. (a) EdgeNet is built on a transformer backbone for edge detection; it outputs the probabilities of 2 classes: Edge / Other. (b) The internal structure of the transformer encoder, featuring a multi-head self-attention mechanism. (c) The schematic of a multi-head attention. (d) ClassNet, implemented using a CNN for multi-class chemical classification. EdgeNet leverages the global contextual modeling capabilities of transformers to accurately detect sample boundaries, while ClassNet utilizes the local feature extraction strengths of CNNs to enable efficient and robust chemical identification within the sample FOV. B: batch size.

ClassNet (Fig. 3d), in contrast, is used for multi-class chemical classification at each pixel and is based on a CNN architecture[54]. It processes individual terahertz pulses through two 1D convolutional layers with ReLU activation and max pooling. The resulting feature maps are flattened and passed through two fully-connected layers. A final softmax layer produces the probability distribution over the target chemical classes per pixel. Together, EdgeNet and ClassNet provide complementary capabilities—enabling accurate spatial boundary detection of the sample under test and robust pixel-level chemical classification within our spectroscopic imaging framework.

We trained both ClassNet and EdgeNet using pharmaceutical and explosive samples. To evaluate their generalization capability, the models were tested on entirely new, previously unseen samples and data. Figure 4 summarizes the blind classification results for these chemical test samples. In Fig. 4a, we present representative images of unknown chemical samples alongside the corresponding ground truth labels, model predictions before the spatial majority voting, and refined predictions after applying spatial majority voting for eight material classes: DCP, MCC, IBU, MAN, $KNO_3$, PETN, RDX, and TNT (see Methods for details). While the raw predictions from the two networks already demonstrated high classification performance, spatial majority voting further improved the results by mitigating spatially isolated pixel-level misclassifications, while also enhancing the edge detection accuracy, correctly revealing the boundaries of chemical objects. The overall pixel-level classification performance of our approach is quantitatively captured in the confusion matrix shown in Fig. 4b, which demonstrates consistent high accuracy across all chemical classes with minimal cross-class confusion. Specifically, the model achieved pixel-level classification accuracies of 99.48%, 98.26%, 100%, 100%, 100%, 99.57%, and 98.89% for DCP, MCC, IBU, MAN, $KNO_3$, PETN, RDX, and TNT, respectively. The average accuracy, including the background region (i.e., the sample holder, without a chemical), reached 99.42%. In addition, Fig. 4c demonstrates our model's strong performance on spatially irregular and cracked test samples, even though it was trained exclusively on intact, regular samples. This successful generalization is rooted in our pulse-based approach. The full time-domain waveforms for intact versus cracked samples differ in pulse timing and relative amplitudes (see Supplementary Fig. S3). However, the characteristic shape of the crucial $P_3$ pulse remains consistent between them. Because our network analyzes individual pulse shapes, it correctly identifies the material despite these structural irregularities. Furthermore, our model demonstrates its generalization capability by accurately predicting four different types of cracked pharmaceutical samples in the same FOV (see Supplementary Fig. S5). These results underscore the robustness of our classification pipeline, highlighting its ability to handle challenging and previously unseen sample conditions.

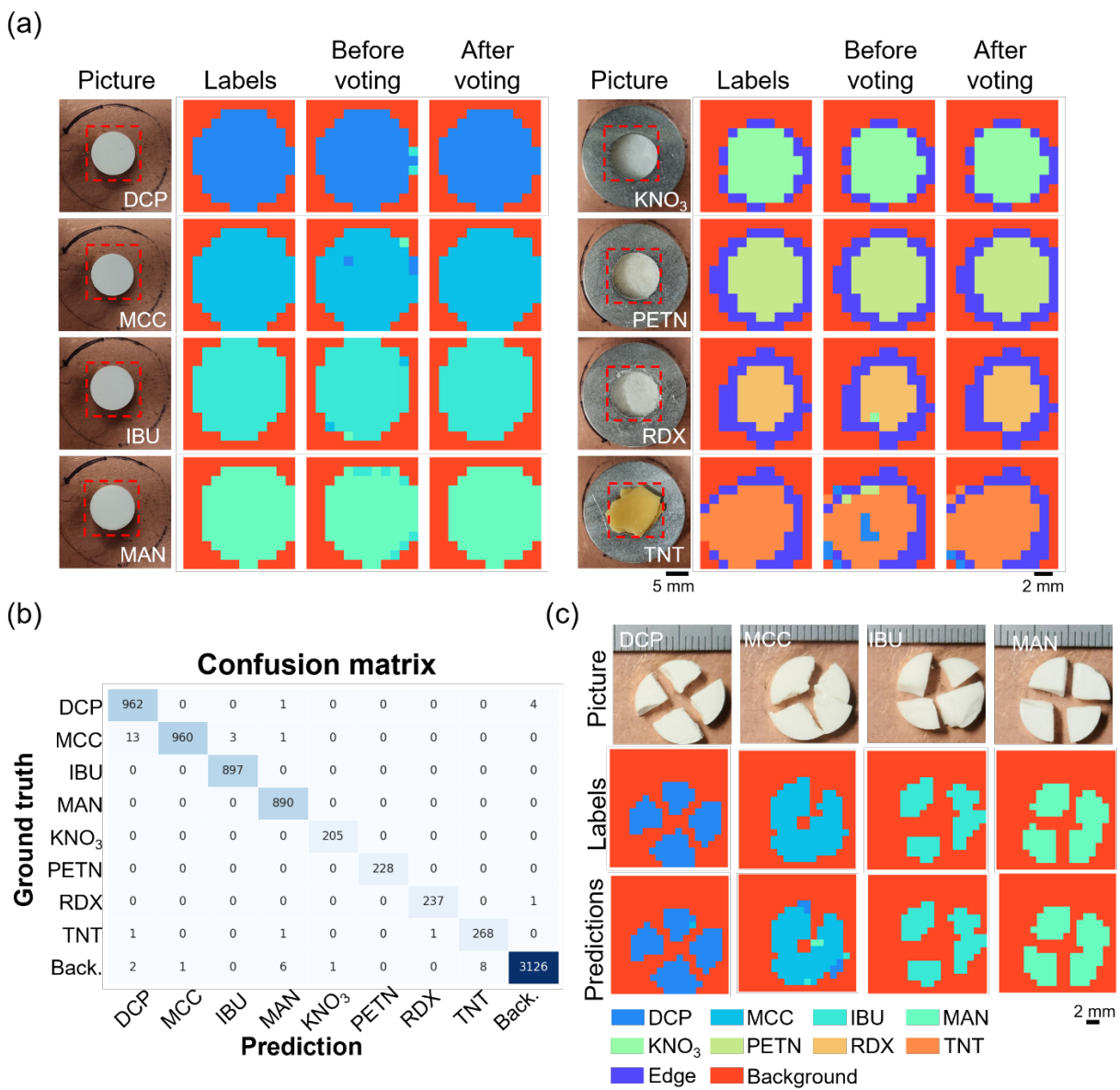

**Figure 4**. Classification results for unobscured chemical samples (pharmaceuticals and explosives). (a) Representative input images, ground truth labels, model predictions before spatial majority voting, and refined predictions after spatial majority voting for DCP, MCC, IBU, MAN, KNO₃, PETN, RDX, and TNT. (b) Confusion matrix summarizing the classification performance across all the chemical classes. (c) Ground truth and model predictions for cracked, irregularly shaped pharmaceutical samples, demonstrating the models' ability to generalize to new test samples despite being trained exclusively on intact, regular samples.

We further evaluated the performance of our classification pipeline on concealed chemical samples, where the substances were hidden beneath a visibly opaque paper cover and not detectable under standard illumination or visible or IR-based machine vision, as illustrated in Fig. 5. Importantly, the network models were trained exclusively on uncovered samples; thus, this evaluation tests their external generalization capability in realistic scenarios where chemical substances may be obscured within envelopes, packages/bags, or other visually inaccessible environments. Figure 5a shows a schematic of the experimental setup, in which each sample is covered/obscured by paper. Representative ground truth labels and our model predictions (after spatial majority voting) for four hidden explosives—$KNO_3$, PETN, RDX, and TNT—are shown in Fig. 5b. Despite the paper-based concealment, our models accurately recover the spatial distributions of these chemicals and specifically identify the type of chemicals.

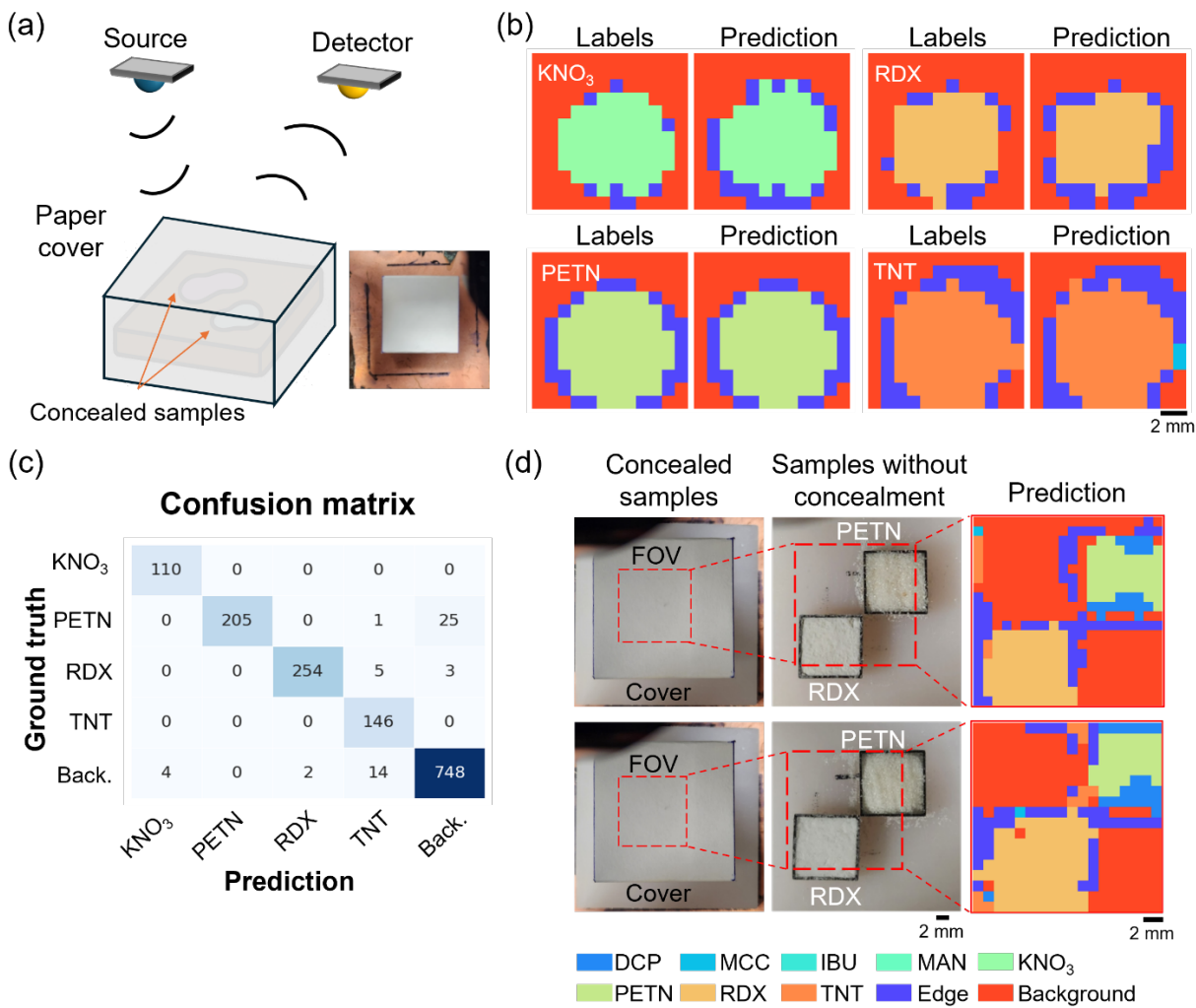

**Figure 5**. Classification results for concealed chemical samples. (a) Schematic illustration of chemical classification through a visibly opaque paper cover; the explosive samples were hidden, i.e., remained undetected under visible illumination. (b) Ground truth labels and model predictions after spatial majority voting for $KNO_3$, PETN, RDX, and TNT. (c) Confusion matrix summarizing the classification performance for the concealed explosive samples. (d) Sample pictures with and without concealment, and our network predictions for a scenario involving multiple types of concealed chemicals within the same field of view. Note that the models were trained exclusively on uncovered samples.

The confusion matrix in Fig. 5c summarizes the classification results, with the model achieving accuracies of 76.20% ($KNO_3$), 94.42% (PETN), 98.64% (RDX), and 87.90% (TNT). The average accuracy, including the background class, is 88.83%. Additionally, Fig. 5d showcases the model's performance in a more complex scenario: identifying multiple types of chemicals concealed under a paper cover within the same FOV. While the cover introduces an additional reflection pulse (see Supplementary Fig. S4), the key pulses of interest retain their characteristic shapes after our extraction process. Consequently, the model successfully differentiates between all hidden chemical regions, highlighting its robustness and generalization under challenging and practically relevant imaging conditions.

**Discussion**

We developed a terahertz time-domain spectroscopic imaging system integrated with deep learning models to enable robust chemical classification of pharmaceutical and explosive materials. A step size of 1 mm was used for raster scanning, which represents a trade-off between the acquisition speed and the classification performance. A finer step size provides spatial redundancy that improves prediction accuracy through majority voting, whereas a larger step size reduces the scanning time at the cost of degraded performance (see Supplementary Fig. S6). We employed a pulse-based analysis method instead of using the full time-dependent spectrum to improve computational efficiency, model robustness, and signal fidelity. The full spectrum is computationally prohibitive due to its large length and temporal sparsity, and the long-range dependencies between reflected pulses are poorly suited for standard 1D CNNs. Our pulse-based analysis strategy effectively decouples chemical identification from confounding physical variables like sample geometry, dimensions, and packaging. We achieve this by individually analyzing each reflected terahertz pulse from the sample. The foundation of this approach is that a material's characteristic pulse shape—which carries the essential chemical information—remains remarkably consistent, even if other factors like sample thickness or packaging change. This enables our network to accurately identify materials under previously unseen conditions. The collected individual pulses are used as independent inputs to train neural networks, which were employed for automated chemical classification and edge detection of unknown samples. Spatial majority voting and morphological operations were applied as post-processing steps to further enhance the classification performance of our chemical imaging system.

Our pipeline achieved high classification accuracy on various chemical samples, reaching an average accuracy of 99.42% across eight chemical classes and the background category. The trained models also demonstrated strong generalization capabilities to previously unseen conditions, such as cracked and spatially irregular samples, despite being trained exclusively on intact ones. Furthermore, we validated the system's external generalization performance by testing it on concealed explosive chemicals hidden beneath opaque paper covers without knowing whether the pulse comes from the sample or the paper cover in the same scanned trace, achieving accurate classification even under these more challenging scenarios. Specifically, we trained the neural networks exclusively on visible samples while reserving concealed cases solely for blind testing. This strategy was adopted to more faithfully represent real-world scenarios, where concealment conditions are variable and often differ from any training set. By structuring the dataset in this way, we were able to rigorously evaluate the external generalization of our models under stringent and realistic conditions.

One limitation of the current study is that the models were trained solely on unconcealed (visible) samples, while the chemical classification of the concealed samples relied entirely on the model's ability to generalize to new types of samples. Although our overall accuracy remained high for such concealed explosives, the performance for $KNO_3$ was notably lower than for other explosives, suggesting that certain chemicals may be more sensitive to concealment-related subtle signal variations. Future studies could explore the trade-off between overall generalization capability and the potential gains in robustness achieved by augmenting the training set with concealed or partially obscured samples, which may improve model performance under more diverse conditions. While our method successfully identifies multiple spatially distinct chemicals in the same FOV, another potential future direction is the quantitative analysis of the concentrations of co-localized mixtures within a single pixel. Future development could focus on a cascaded framework to first detect the presence of a mixture and then estimate constituent concentrations of each chemical that is identified. Additionally, optimizing network architectures or incorporating self-supervised pretraining strategies may reduce the dependence on large labeled datasets. Expanding the diversity of the training dataset—by adding more chemical types, irregular sample geometries, or varied concealment/packaging materials—would further strengthen the classification accuracy of the proposed system.

Our pulse-based terahertz imaging method provides a powerful tool for obtaining critical chemical intelligence in a non-contact manner. While data acquisition for a 12x12 mm² area takes approximately ~7.5 minutes (~447 seconds), the subsequent post-processing and pixel-level classification by the neural network are completed in ~1 second. This capability for non-destructive imaging followed by near-instantaneous digital analysis is ideal for applications in e.g., pharmaceutical quality control and security screening. The system enables the non-destructive analysis of individual tablets and capsules, potentially allowing for the verification of active ingredients and the detection of counterfeit drugs. In the security and forensic domain, it could be used for identifying trace explosives or illicit substances collected on swabs and for inspecting suspicious powders or stains in mail. This provides a powerful tool for obtaining critical chemical intelligence in a non-contact manner.

While the proof-of-concept experiments presented here rely on mechanical raster scanning to capture 2D chemical images, this requirement can be eliminated by replacing the single-pixel terahertz detector with a terahertz focal-plane array based on plasmonic nanoantenna arrays[55,56]. This advancement could enable significantly faster imaging and reduce the system's complexity, size, and weight, while preserving the high signal-to-noise ratio and broad bandwidth necessary for accurate spectroscopic analysis and chemical imaging.

Overall, this study demonstrates a powerful framework for terahertz-based chemical imaging, offering a promising pathway toward robust, high-accuracy classification of chemicals in both controlled and operational environments where concealment, packaging, and sample variability pose major challenges. This framework holds strong potential for various applications in e.g., security screening, pharmaceutical quality control, and non-destructive testing, where rapid and reliable material identification is critical.

**Materials and methods**

**Terahertz time-domain spectroscopy system**

A mode-locked Ti:sapphire laser (Coherent Mira 900) was used to generate an optical pulse train with a repetition rate of 76 MHz, a central wavelength of 800 nm, and a pulse width of 135 fs. The laser output was split into pump and probe beams using an optical beam splitter. The femtosecond optical pump beam illuminates the photoconductive antenna, generating photocarriers, where the optical absorption is further enhanced by the custom-designed plasmonic structure. An external bias voltage is applied to accelerate these carriers, producing an ultrafast photocurrent that radiates broadband terahertz pulses. The bias voltage is modulated at 100 kHz to enable lock-in detection. The probe beam was routed through an optical delay line (Newport DL125) to a plasmonic photoconductive terahertz detector. The optical powers incident on the emitter and detector were 600 mW and 150 mW, respectively. Terahertz radiation emitted from the source was collimated using a gold-coated off-axis parabolic mirror and subsequently focused onto the sample using a second gold-coated off-axis parabolic mirror. The focused terahertz beam had a spot diameter of ~2 mm at the sample location. The sample was mounted on a copper platform attached to a motorized stage, allowing raster scanning with a step size of 1 mm. Reflected terahertz radiation from the sample was redirected using a 50/50 terahertz beam splitter and focused onto the detector via an additional off-axis parabolic mirror. Similar to terahertz generation, the femtosecond probe beam generates photocarriers in the active area of the photoconductive terahertz detector. The electric field of the incident terahertz pulse acts as a transient bias, accelerating these photocarriers to the antenna electrodes. This acceleration induces a photocurrent that is directly proportional to the terahertz electric field's amplitude at the precise instant the probe beam arrives. By varying the optical path length between the two branches with the delay line, the time-domain terahertz electric field was retrieved by measuring the photocurrent generated in the terahertz detector. This photocurrent was converted to a voltage and amplified using a transimpedance amplifier (DLPCA-200, FEMTO). The output voltage was then fed into a digital lock-in amplifier (MFLI, Zurich Instruments), and the resulting data were transferred to a computer for post-processing and analysis.

**Preparation of chemical samples**

The four pharmaceutical materials used in this study were microcrystalline cellulose (VIVAPUR 200), dibasic calcium phosphate dihydrate (EMCOMPRESS), mannitol (PEARLITOL 100 SD), and ibuprofen (DC 100). Each pharmaceutical powder was compressed into a cylindrical tablet using a hydraulic press. The resulting tablets had a diameter of 10 mm and a thickness ranging from 2 mm to 4 mm. The four explosives examined were potassium nitrate (Lab Alley, powder, USP/ACS/FCC grade), pentaerythritol tetranitrate (Dyno Nobel Inc., R24010), cyclotrimethylene trinitramine (Dyno Nobel Inc., ME006), and trinitrotoluene (Dyno Nobel Inc., ME001CA). These materials were placed in a custom-fabricated aluminum mount featuring a central cylindrical hole with a diameter of 10 mm and a depth of 5 mm. For measurements involving concealed explosives, the samples were placed on a polyethylene sheet and enclosed within a plastic square frame to enable the placement of two samples within the FOV. The square frame, fabricated via 3D printing, had a width of 10 mm and a depth of 3 mm. A paper cover with a thickness of 0.18 mm was shaped into a rectangular box with a lateral size of 35 mm × 35 mm and a height of 10 mm to conceal the samples during testing.

**Pulse extraction and training data generation**

We automatically extract individual pulses from each time-domain waveform using a peak detection function configured with an 8% amplitude threshold (relative to a reference reflection from the background region) and a minimum separation of 650 data points, which corresponds to 13 ps. After the pulse extraction step, we perform a semi-automated labeling process for network training and label creation. We first identify the background-reflected pulse ($P_2$) by its consistent arrival time across the scanned area and label it as "background". All other detected pulses (e.g., $P_1$, $P_3$, $P_4$) are assigned a specific chemical label based on a priori knowledge of the sample being scanned. Concurrently, a second, sample-specific amplitude threshold is used to re-label any pixel with only low-amplitude signals as an "edge". The final output is a training dataset composed of these labeled, isolated pulses, which are collectively used to train and validate the neural network for pixel-level classification of chemical types, edges, and background.

**Integrated chemical classification pipeline and spatial post-processing of chemical images**

EdgeNet identifies edge pixels by analyzing spatial boundary information across all the detected pulses (per pixel), whereas ClassNet operates on individual pulses to generate per-pixel chemical class probabilities. The outputs from EdgeNet and ClassNet are integrated via a conditional decision rule. Specifically, pixels identified by EdgeNet as potential edges and simultaneously exhibiting a maximum class probability (assigned by ClassNet) below a predefined threshold (empirically set to 0.95) are classified as "Edge." All other pixels retain their original labels assigned by ClassNet. This integrated inference procedure is applied iteratively across the entire sample FOV to generate an initial chemical classification map. To reduce potential misclassifications and further enhance inference robustness, a post-processing step is performed that incorporates spatial majority voting and morphological filtering. Majority voting is applied within a 3×3 neighborhood of each pixel to enforce local consistency. Morphological operations based on eight connected pixel structures are then used to eliminate spatially isolated errors and remove spurious regions. This refinement step yields spatially consistent classification results, producing chemical images that more accurately represent the true spatial distribution and identity of the target substances across the sample FOV.

**Model training setup**

EdgeNet and ClassNet were trained separately using Python (v3.8.16) and PyTorch (v1.11, Meta AI) on a workstation equipped with a GeForce RTX 3090 GPU (Nvidia Corp.), an Intel® Core™ i9-12900KF CPU (Intel Corp.), and 64 GB of RAM, running Windows 11 (Microsoft Corp.). Each model was independently optimized using the Adam optimizer[57] with a learning rate of $1 \times 10^{-3}$. Training was performed over 200 epochs with a batch size of 64, requiring approximately 1.5 hours for ClassNet and 2.5 hours for EdgeNet.

8. Kemp, M. C. Explosives detection by terahertz spectroscopy - A bridge too far? *IEEE Trans. Terahertz Sci. Technol.* **1**, 282–292 (2011).

9. Puc, U. et al. Terahertz spectroscopic identification of explosive and drug simulants concealed by various hiding techniques. *Appl. Opt.* **54**, 4495-4502 (2015).

10. Palka, N. THz reflection spectroscopy of explosives measured by time domain spectroscopy. *Acta Physica Polonica A* **120**, 713-715 (2011).

11. Alexander, N. E. et al. TeraSCREEN: multi-frequency multi-mode Terahertz screening for border checks. Proc. SPIE 9078, Passive and Active Millimeter-Wave Imaging XVII, 907802 (Baltimore, 2014).

12. Davies, A. G., Burnett, A. D., Fan, W., Linfield, E. H. & Cunningham, J. E. Terahertz spectroscopy of explosives and drugs. *Materials Today* **11**, 18–26 (2008).

13. Trofimov, V. A. & Varentsova, S. A. Efficiency of using the spectral dynamics analysis for pulsed THz spectroscopy of both explosive and other materials. Proc. SPIE 9454, Detection and Sensing of Mines, Explosive Objects, and Obscured Targets XX, 945409 (Baltimore, 2015).

14. Chen, Y., Ma, Y., Lu, Z., Qiu, L. & He, J. Terahertz spectroscopic uncertainty analysis for explosive mixture components determination using multi-objective micro-genetic algorithm. *Advances in Engineering Software* **42**, 649–659 (2011).

15. Moradikouchi, A. et al. Terahertz frequency-domain sensing combined with quantitative multivariate analysis for pharmaceutical tablet inspection. *Int. J. Pharm.* **632**, 122545 (2023).
28

**Acknowledgements**

The authors acknowledge funding from the DARPA FLEX Program Award # HR0011-24-2-0353. Development of the plasmonic nanoantenna arrays used in this work was supported by the Department of Energy (grant # DE-SC0016925).

# Supplementary Information for

# Detection and imaging of chemicals and hidden explosives using terahertz time-domain spectroscopy and deep learning

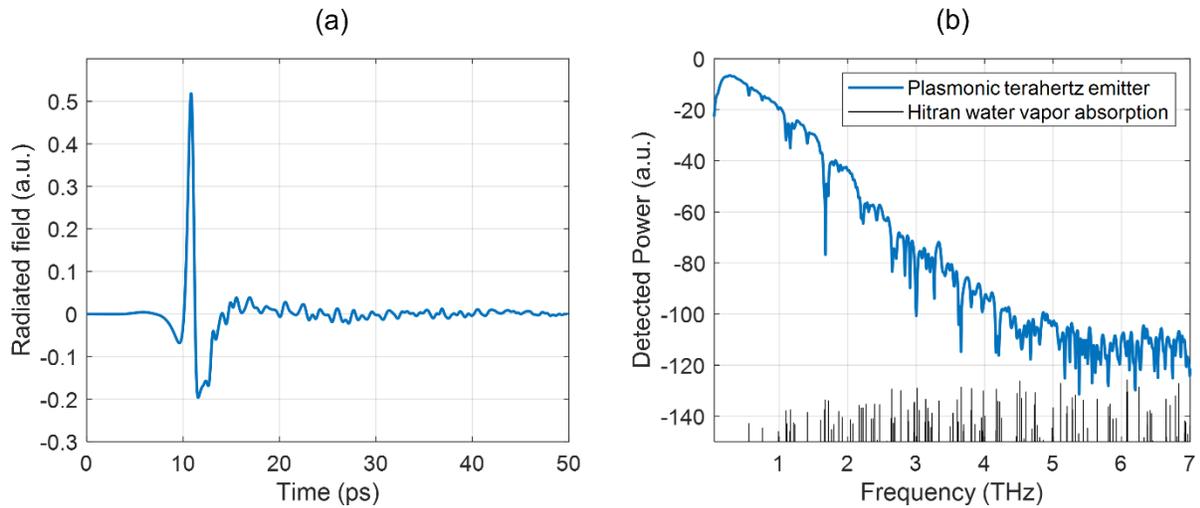

**Figure S1.** (a) Terahertz time-domain response of the developed THz-TDS system in transmission mode. (b) Corresponding power spectrum. To benchmark against state-of-the-art systems—typically characterized in transmission mode—we evaluated our system in transmission mode. The system achieves a peak dynamic range of 104 dB and a bandwidth of 5 THz for a 50 ps scan range and a 3-second scan time. Note that in reflection mode, the use of a 50:50 terahertz beam splitter introduces at least a 6 dB reduction in signal-to-noise ratio (SNR).

| THz-TDS system | Dynamic range (dB) | Bandwidth (THz) | Measurement time (s) | Scan range (ps) |
|---|---|---|---|---|
| Ours | 104 | 5 | 3 | 50 |
| Menlo Tera K15[1] | 110 | 6 | 60 | 50 |
| Menlo TeraSmart[2] | 100 | 6.5 | 5 | 50 |
| TeraFlash pro[3] | 95 | 6 | 20 | 50 |
| TeraFlash smart[4] | 80 | 4.5 | 60 | 150 |
| Advantest TAS7500TS[5] | 70 | 4 | 262 | 131 |

**Table S1.** Performance comparison of our THz-TDS system in transmission mode with the state-of-the-art systems.

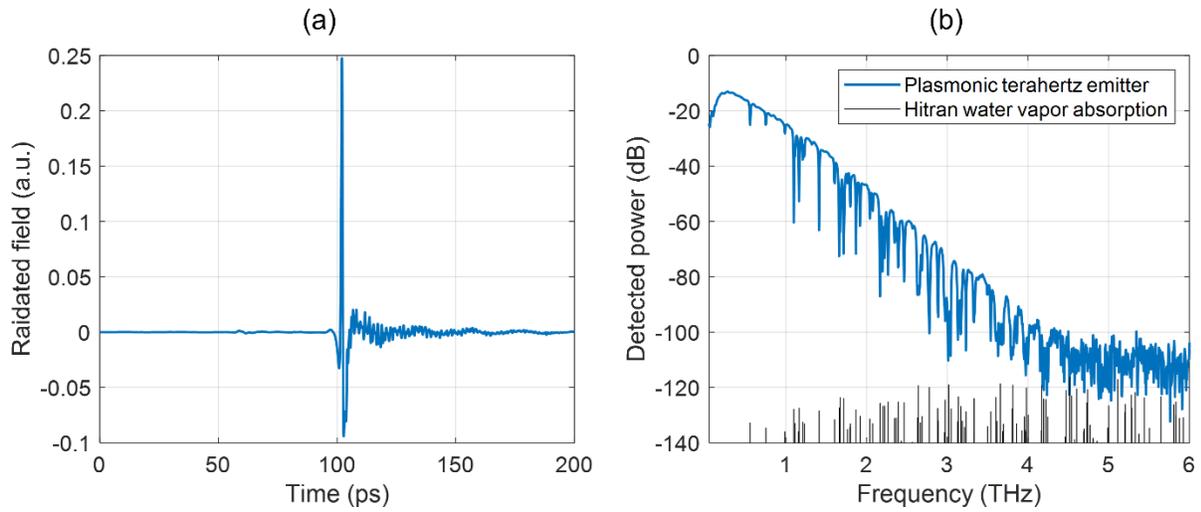

**Figure S2.** (a) Terahertz time-domain response of the system without a sample on the copper platform. (b) Corresponding terahertz power spectrum. The system achieves a peak dynamic range of 96 dB and a bandwidth of 4.5 THz in a measurement time of 3 seconds.

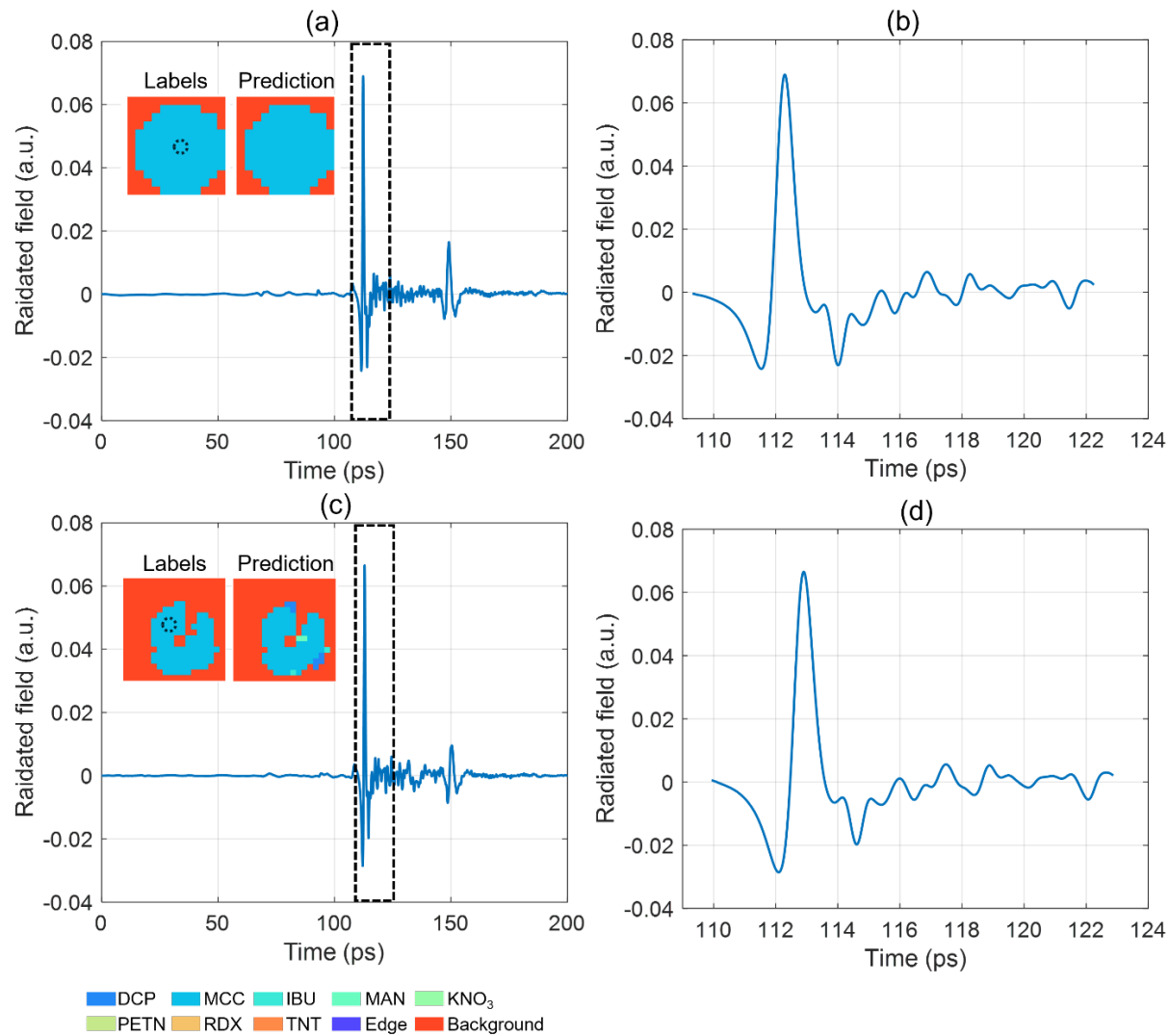

**Figure S3. Terahertz time-domain responses of MCC.** (a) The full time-domain response of an intact MCC tablet at one pixel. Inset figures: the scanned pixel in the labeled ground truth and the prediction of the network. (b) One of the separated pulses from (a), which is P3, i.e., the pulse reflected from the bottom of the tablet. (c) The full time-domain response of a cracked MCC tablet at one pixel. Inset figures: the scanned pixel in the labeled ground truth and the prediction of the network. (d) One of the separated pulses from (c), which is P3, i.e., the pulse reflected from the bottom of the tablet. The time delay and relative amplitude between two pulses are different in (a) and (c). The separated pulses in (b) and (d) are similar, enabling accurate pulse-based predictions.

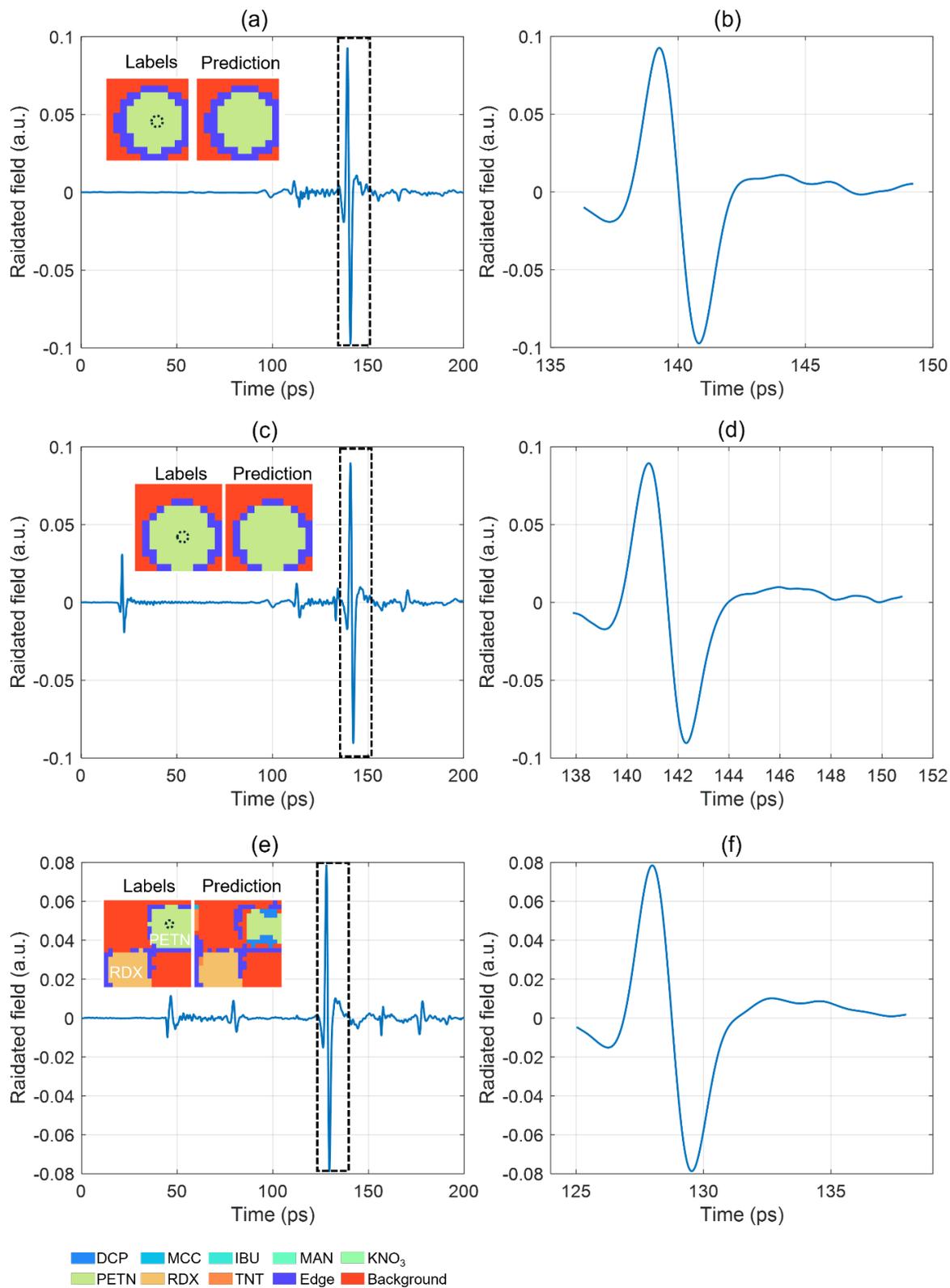

**Figure S4. Terahertz time-domain responses of PETN.** (a) The full time-domain response of PETN at one pixel. Inset figures: the scanned pixel in the labeled ground truth and the prediction of the network. (b) One of the separated

pulses from (a), which is P3, i.e., the pulse reflected from the bottom of PETN. (c) The full time-domain response of PETN at one pixel under the paper cover. The first pulse in the trace is the reflection from the paper cover. Inset figures: the scanned pixel in the labeled ground truth and the prediction of the network. (d) One of the separated pulses from (c), which is P3, i.e., the pulse reflected from the bottom of PETN. (e) The full time-domain response of PETN at one pixel under a paper cover with RDX in the same FOV. (f) One of the separated pulses from (e). Although (a), (c), (e) are different because of geometric differences, the separated pulses (b), (d), (f) are similar. The neural network analyzed all the pulses in (a), (c), (e) individually without knowing whether the pulse came from the paper cover or the sample, achieving accurate detection.

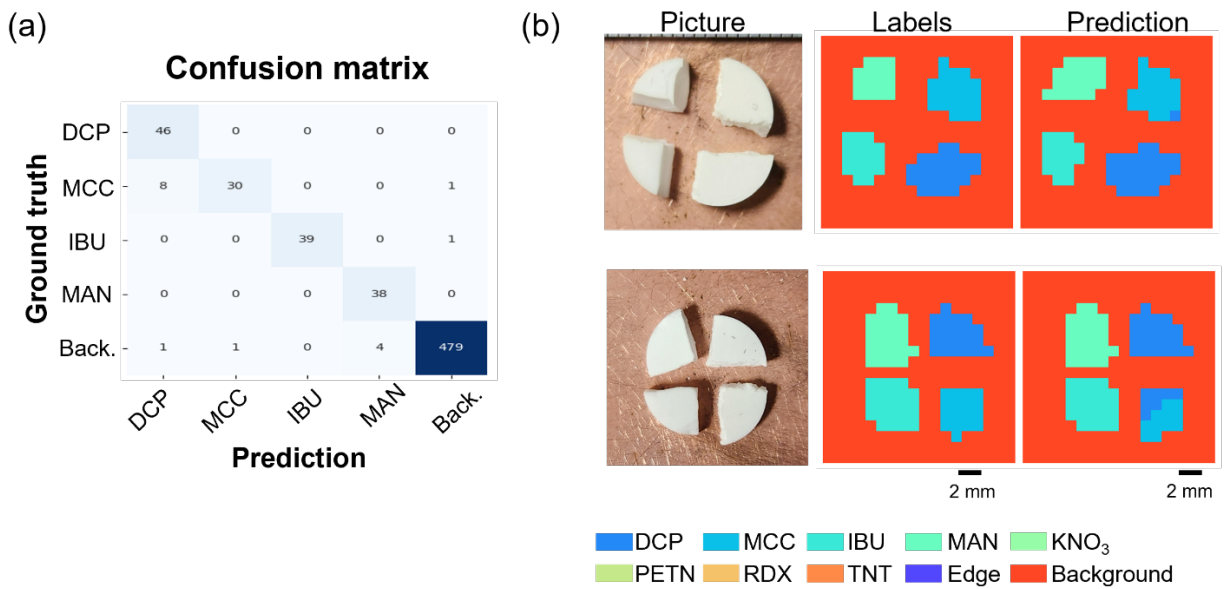

**Figure S5. Classification results for four different types of cracked pharmaceutical samples in the same field of view.** (a) Confusion matrix summarizing the classification performance. The classification accuracies for DCP, MCC, IBU, MAN, and background are 100%, 82.86%, 100%, 100%, and 98.76%, respectively. The average accuracy is 98.13%. (b) Pictures of the cracked samples, ground truth labels and our network predictions.

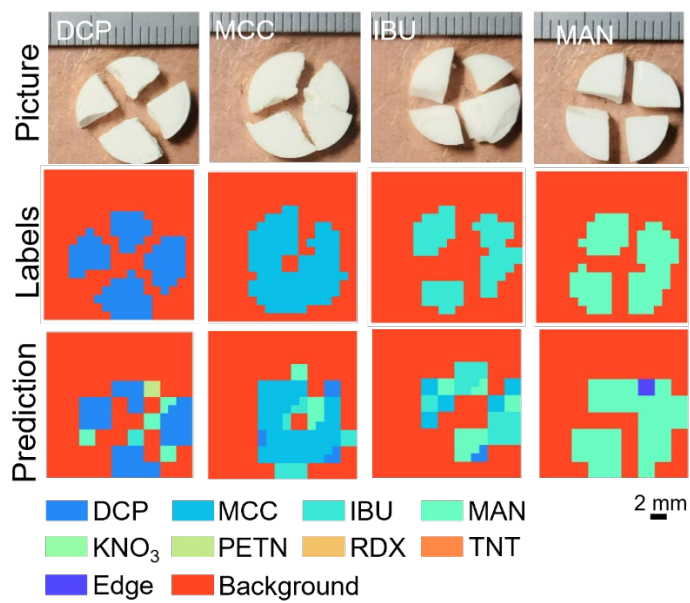

**Figure S6. Classification results for cracked pharmaceutical samples with a step size of 2 mm.** To investigate the effect of the spatial resolution, we down-sampled the dataset for the cracked samples (Fig. 4c) to simulate a coarser 2-mm step size. Compared to the 1-mm results, this led to a significant degradation in performance, with prediction accuracies dropping to 67.61% for DCP, 60.93% for MCC, 51.30% for IBU, 82.01% for MAN, and 91.43% for the background. These results underscore that fine spatial resolution is crucial for accurately classifying irregularly shaped chemical samples. A higher resolution provides more pixels for a given area, which enhances the reliability of the majority voting process and, in turn, boosts the final prediction accuracy.